\documentclass [aps,twocolumn,groupeauthor,showpacs,superscriptaddress,prl]{revtex4-1} 
\pdfoutput=1
\usepackage{amsmath}
\usepackage{amssymb}
\usepackage{bm}

\usepackage{graphicx}
\usepackage{color}
\usepackage{soul}

\newcommand{\ab}{{\alpha\beta}}

\begin{document}
\title{Nonequilibrium Green's functions for functional connectivity in the brain}

\begin{abstract}
A theoretical framework describing the set of interactions between neurons in the brain, or functional connectivity, should include dynamical functions representing the propagation of signal from one neuron to another. Green's functions and response functions are natural candidates for this but, while they are conceptually very useful, they are usually defined only for linear time-translationally invariant systems. The brain, instead, behaves nonlinearly and in a time-dependent way. Here, we use nonequilibrium Green's functions to describe the time-dependent functional connectivity of a continuous-variable network of neurons. We show how the connectivity is related to the measurable response functions, and provide two illustrative examples via numerical calculations, inspired from \textit{C. elegans}.
\end{abstract}
\author{Francesco Randi}
\email[Corresponding author: ]{frandi@princeton.edu}
\affiliation{Department of Physics, Princeton University, Jadwin Hall, Princeton, NJ 08544, USA}
\author{Andrew M. Leifer}
\email[Corresponding author: ]{leifer@princeton.edu}
\affiliation{Department of Physics, Princeton University, Jadwin Hall, Princeton, NJ 08544, USA}
\affiliation{Princeton Neuroscience Institute, Princeton University, NJ 08544, USA}
\date{\today}
\maketitle

Understanding how neurons interact is fundamental to describing how their collective activity generates the complex dynamics of the brain. Advances in optogenetics and neuroimaging now allow activity to be stimulated in one neuron while simultaneously measuring the response of many others in a network~\cite{Rickgauer2014,Emiliani2015,Yang2018}, providing insights into how signals travel through the brain. Functional connectivity encompasses the collection of strengths, signs, and time-varying properties that govern how a change in activity of one neuron affects another. Measuring functional connectivity would constrain simulations by providing a missing link between the anatomical connectivity and the neural dynamics. Further, measuring how functional connectivity changes can reveal what dynamical properties of the brain change with learning.

Most existing models of continuous-variable neural activity are formulated as differential equations~\cite{ErmentroutBook2010,DayanAbbotBook2001,Wicks1996,Kunert2014,Kunert2017a,Kunert2017b}. These equations include parameters for local properties of direct connections in the network, such as the strengths of the synapses between two neurons. But those local properties cannot be measured directly in the network. Instead, experiments see an effective interaction between the two neurons, which includes contributions from indirect paths as well as the direct path.

An integral formulation, such as \cite{Brinkman2018}, is a more convenient formalism for transitioning between local direct connections and the effective ones that are more experimentally accessible. In the linear and time-translationally invariant (TTI) case (a condition that we will relax in this paper), the activity $\psi_i$ of neuron $i$ is 
\begin{equation}
\label{eq:lin_time_inv}
\psi_i(t) = \psi_{i,\mathrm{eq}} + \sum_{j\in \mathrm{all}}g_{0,ij} * \Delta\psi_j\, (t) + g_{0,i}^\mathrm{ext}*I^\mathrm{ext}_i(t),
\end{equation}
where $\psi_\mathrm{eq}$ are the equilibrium activities of the neurons (which depend on the rest of the network) and $\Delta\psi$ the deviations from those values. $*$ denotes a convolution, $g_{0,ij}$ is the (TTI) Green's function, or transfer function, describing the direct interaction $i\leftarrow j$ from neuron $j$ to neuron $i$. 
$g_{0,i}^\mathrm{ext}*I^\mathrm{ext}_i(t)$ denotes the effect of external perturbations.

Eq.~\eqref{eq:lin_time_inv} considers only direct paths between neurons. However, in a network 
$i$ and $j$ are connected by both direct and indirect paths,
and one would have to solve Eq.~\eqref{eq:lin_time_inv} for each neuron and each time-step. 
If we know $\Delta\psi_j$~\cite{footnotem1}
and want to calculate $\Delta\psi_i$, in a linear system we can condense the effect of the whole network in a single connected Green's function $G^{j}_{0,ij}$ (the resolvent kernel in Volterra integral equations~\cite{Linz1985}), such that  $\psi_i(t) = \psi_{i,\mathrm{eq}} + (G^{j}_{0,ij} * \Delta\psi_j) (t)$. $G^{j}_{0,ij}$ is a solution to
\begin{equation}
\label{eq:fullyconn}
G^{j}_{0,ij} = g_{0,ij} + \sum_{\mu\neq j}g_{0,i\mu} * G^{j}_{0,\mu j},
\end{equation}
which is obtained recursively inserting the contributions of all the neurons in Eq.~\eqref{eq:lin_time_inv} (upper case is for connected, subscript 0 for linear and TTI, superscript $j$ means that $j$ is excluded from the sums. For when $\Delta\psi_j$ is sufficient and for the derivation, see the Supplement~\cite{Supplement}).
To probe the system, we can induce a perturbation $\delta\psi_j$ on top of the current state of the system $\bm{\psi=\psi_{\mathrm{eq}}+\Delta\psi}$ and obtain the connected response function $F_{0,ij}$ by measuring the produced $\delta\psi_i = F_{0,ij}*\delta\psi_j$. In the linear and TTI case, $F_{0,ij}=G^{j}_{0,ij}$.

One reason Green's functions have found only limited use in neuroscience~\cite{DayanAbbotBook2001,Brinkman2018} is that Green's functions are usually defined only for linear and TTI systems, while the brain is highly nonlinear. Nonlinearities allow the brain to perform nontrivial computations and to have responses that depend on past history or sensory context. Nonlinear corrections to a Green's function-like formulation via systematic expansion has previously been used to describe the effect of hidden neurons~\cite{Brinkman2018} and spike train statistics~\cite{Ocker2017}. Because the concept of a response function is intuitive, and an experiment can always be designed to measure a response function, it is worth working with expanded, or corrected, Green's functions. 

In this work we use nonequilibrium Green's functions $G_{ij}$~\cite{NEDMFT,Ocker2017,HerreraDelgado2020} to describe the time-dependent functional connectivity of a continuous-variable network of neurons, and discuss their relation to the nonequilibrium response functions $F_{ij}$ measured in experiments (absence of subscript 0 means nonequilibrium). 
While they retain the benefits of transfer functions, their nonequilibrium definition as a function of relative \emph{and} absolute time makes them well-suited to capture nonlinearities and time-dependence in the brain, for example when synapses saturate, when synaptic adaptation occurs, or when neuromodulators change the cellular properties of the neurons in a time-dependent way. Nonequilibrium Green's functions are used in other fields, like the theory of many-body systems in condensed matter physics, where they guide both theory and experiments~\cite{NEDMFT}. Note that here equilibrium refers to the time-invariance of the Green's functions, not the neural activities. 

We first present a general model-independent equation for the connected nonequilibrium response functions $F_{ij}$ (Eq.~\eqref{eq:fullyint}), that allow us to write $\delta\psi_i(t) = F_{ij}*\delta\psi_j = \int dt_1\,F_{ij}(t,t_1)\delta\psi_j(t_1)$, and are obtained assuming sparse nonlinear connections, or edges, $(\alpha,\beta)$. These edges are described with nonequilibrium Green's functions $g_{\ab}[\bm{\psi}]$ so that, formally, $\Delta\psi_\alpha=g_{\ab}[\bm{\psi}]*\Delta\psi_\beta=\int dt_1\,g_{\ab}[\bm{\psi}](t,t_1)\Delta\psi_j(t_1)$ for an isolated pair of neurons. Because $g_{\ab}[\bm{\psi}]$ is functionally dependent on the state $\bm{\psi}$ of the system it has to be calculated according to its nonlinear expression. Once it is calculated, however, other properties of the network, like the other Green's functions and the response functions, are easily derived and computed.

We will describe how the $F_{ij}$s relate functional connectivity to experiments, apply this formalism to the nervous system of the nematode worm \textit{Caenorhabditis elegans}, and illustrate the general theoretical results with numerical calculations.

\paragraph{Nonequilibrium response functions}
As we derive an equation for the nonequilibrium response function $F_{ij}$, we will also address a seemingly puzzling experimental observation about the \textit{C. elegans} nervous system.
Characterizations of some synpases in the worm have shown that they are linear throughout a large part of the physiological range of membrane potentials~\cite{Liu2009,Lindsay2011,Narayan2011}. However, we know that nonlinearities and time-dependence are critically important in the \emph{C. elegans} nervous system and in nervous systems generally, because they allow the network to perform  computations, including for example responding  to sensory stimuli in a context dependent manner~\cite{Mochi2019,Dobosiewicz2019}. How does a network have many linear edges but also show widespread nonlinear behaviors? In the integral formulation with nonequilibrium Green's functions it is straightforward to show how these two observations can coexist. 

We start by considering a network in which only one of the edges, $(\alpha,\beta)$, displays a significant nonlinearity. This is in contrast to an approach in \cite{Ocker2017} which assumes nonlinearities that are homogeneous over the network and then proceeds with their systematic expansion.  We will show how a time-dependent change of a single edge, due e.g. to a nonlinearity, can change effective connections and response functions elsewhere in the network.

The direct Green's function $g_\ab(t,t')=g_{0,\ab}(t-t')+\pi_\ab[\bm{\psi}](t,t')$ for the nonlinear or time-dependent edge can be written as the sum of a linear and TTI term $g_{0,\ab}$ and a nonequilibrium term $\pi[\bm{\psi}]$, which depends on the state $\bm{\psi}$ of the system. For the isolated pair $\alpha\leftarrow\beta$, $g_\ab$ allows one to calculate the response function $f_\ab$ that determines $\delta\psi_\alpha$ measured in an experiment after a perturbation $\delta\psi_\beta$ on top of the current state $\bm{\psi=\psi_{\mathrm{eq}}+\Delta\psi}$. With $f_\ab$, one can write $\delta\psi_\alpha=f_\ab*\delta\psi_\alpha=\int dt_1\,f_\ab(t,t_1)\delta\psi_\beta(t_1)$, where nonlinearities and time-dependence are implicitly taken into account in the nonequilibrium $f_\ab$,
\begin{align}
\label{eq:nonlinedge}
f_\ab(t,t') &= g_{0,\ab}(t-t') + \bar\chi_\ab[\bm{\psi}](t,t'),\\
\label{eq:pi}
\bar\chi_\ab[\bm{\psi}](t,t') &= \pi_\ab(t,t') + \biggl(\frac{\delta\pi_\ab(t,t_1)}{\delta\psi_\beta(t')}*\psi_\beta(t_1)\biggr)(t,t').
\end{align}
(see Supplemental material~\cite{Supplement} for more details)

The connected nonequilibrium response function $F_{ij}$ of a general effective edge $(i,j)$ in a network is obtained following similar steps to the ones leading to Eq.~\eqref{eq:fullyconn}, but using Eq.~\eqref{eq:nonlinedge} for the edge $(\alpha,\beta)$, and is
\begin{equation}
\label{eq:fullyint}
F_{ij}(t,t') = F_{0,ij}(t-t') + \big(F^j_{0,i\alpha}* \bar\chi_\ab* F_{\beta j}\big)(t,t'),
\end{equation}
where $(A*B)(t,t')=\int dt_1\,A(t,t_1)B(t_1,t')$. The response $\delta\psi_i$ to a perturbation $\delta\psi_j$ can be written as a simple convolution $\delta\psi_i(t) = (F_{ij}*\delta\psi_j)(t)$ where $F_{ij}$ evolves due to the nonequilibrium terms $\bar\chi[\bm{\psi}]$ (and $\pi[\bm{\psi}]$). 
The $\pi[\bm{\psi}]$ and $\bar\chi[\bm{\psi}]$ we will consider below can be derived exactly, but there is no one recipe for calculating all possible $\pi[\bm{\psi}]$ and $\bar\chi[\bm{\psi}]$. Condensed matter physics provides useful approximations and techniques for calculating them in more complicated cases~\cite{NEDMFT}. We use the notation $F_0$ instead of $G_0$ to emphasize that we are discussing a response function. Eq.~\eqref{eq:fullyint} contains different terms (see Supplemental material~\cite{Supplement}) when $i$ and/or $j$ are equal to $\alpha$ and $\beta$.

The more the neurons on the edge $(\alpha,\beta)$ act as hubs in the network, the larger the fraction of the functional connectivity is affected by their nonlinearity. For example, $\beta$ could be an interneuron integrating inputs from multiple neurons.
Sensory neurons can also act as hubs. Increasing evidence shows that, in \textit{C. elegans}, sensory neurons are well interconnected with the rest of the network~\cite{Dobosiewicz2019}. The application of a sensory stimulus could drive $(\alpha,\beta)$ in a nonlinear regime and, therefore, alter effective interactions between other neurons. The existence of many types of hubs in neural networks make the framework presented here particularly valuable.

Eqs.~\eqref{eq:nonlinedge} and~\eqref{eq:fullyint} describe an approximately linear regime on top of an arbitrary state of the system. Switching linear dynamical systems (SLDS) models~\cite{Linderman2019,Costa2019} assume that such locally linear regimes exist. They describe the dynamics of a nonlinear system as a temporal sequence of linear systems each with different parameters, and have previously been applied to \textit{C. elegans}. In existing SLDS models, the time-dependent switching between parameters is entirely phenomenological. In our approach, Eqs.~\eqref{eq:nonlinedge},~\eqref{eq:fullyint}, and supplementary Eq.(14)~\cite{Supplement}, explicitly govern how nonlinearities in the network produce time-dependent changes to a linear system. Here the response functions contain the time-dependent parameters of the SLDS.

This framework has both computational and conceptual advantages. 
Once the nonlinear $\pi_\ab[\bm{\psi}]$ is calculated, the $F_{ij}$ can be calculated for a given effective edge $(i,j)$ via simple convolutions and without needing all the details of the network. In fact, it is only necessary to run the calculation for two effective edges: the selected edge $(i,j)$ and $(\beta,j)$. If the network has significant nonlinearities on multiple edges, the approach can still be used to calculate response functions, except now the last term in Eq.~\eqref{eq:fullyint} becomes a summation running over all the nonlinear edges $(\alpha,\beta)$, and therefore the nonlinear calculation becomes more computationally intensive. 

\paragraph{Experimental characterization}
Importantly, $F_{ij}$ are the response functions that can be obtained in experiments on networks of neuron as responses to impulsive perturbations. The $F_{ij}$s are always well defined experimentally and theoretically, whether one is studying a complete or subsampled network (see the Experimental characterization section in the Supplemental~\cite{Supplement}).

The local $f_{ij}$ are also of interest, however, because they are directly related to the anatomical connections between the neurons and to the molecular mechanisms responsible for the interactions between them. For models that use equations in differential form, several approaches have been proposed to fit local parameters from spontaneous neural activity, especially in spiking neurons~\cite{Pillow2007,Soudry2015,Dunn2007,Tyrcha2014,Bravi2017}.

In the integral formulation, to obtain the local $f_{ij}$ from the measured $F_{ij}$, one can use deconvolutions and equations~\eqref{eq:fullyconn} and~\eqref{eq:fullyint} under the condition of having a complete measurement of $F_{ij}$ for each pair $(i,j)$ and a suitable ``scan'' across the nonlinearities. While this is experimentally impractical for larger animals, it might be achievable soon on smaller ones like \textit{C. elegans}. However, (de)convolutions are particularly susceptible to noise, so that the response functions might need to be parametrized depending on the level of noise. The ability to selectively introduce nonlinearities and the availability of fast routines to calculate the response functions will prove very valuable in fits, where functions have to be evaluated several times. We leave to future work the details of obtaining $f_{ij}$ from $F_{ij}$.

\paragraph{C. elegans nervous system}
To do calculations, we need to provide explicit expressions for the equilibrium Green's function and the nonlinear term $\pi$, beyond Eq.~\eqref{eq:fullyint}. We consider the equations used in~\cite{Wicks1996,Kunert2014} to simulate neural dynamics in \textit{C. elegans}. Here, the neural activity $\psi_i$ is the membrane potential $V_i$, and each neuron $i$ is described as a single electrical compartment~\cite{Wicks1996,Kunert2014} via the equation
\begin{equation}
\label{eq:kunertv}
\partial_t V_i = -\gamma_{i}(V_i - E_{c,i}) - \gamma_{ij}^{\mathrm{g}}(V_i-V_j) - \gamma_{ij}^{\mathrm{s}}s_{ij} (V_i-E_{ij}),
\end{equation}
where the constants $\gamma$ have dimensions of a conductance over a capacitance and describe leakage ($\gamma_{i}$); electrical synapses, or gap junctions ($\gamma_{ij}^\mathrm{g}$); and chemical synapses ($\gamma_{ij}^\mathrm{s}$). $E_{c,i}$ is the reversal potential of the leaking channels, and $E_{ij}$ the reversal potential of the ionotropic receptors at the synapse. $s_{ij}$ is a synaptic activity variable that evolves according to
\begin{equation}
\label{eq:kunerts}
\partial_t s_{ij} = a_r \phi_{ij}(V_j) (1-s_{ij}) - a_d s_{ij}.
\end{equation}
$\phi_{ij}(V_j)$ describes the dependence of the calcium influx in the presynaptic site on the presynaptic voltage, which triggers the release of vesicles into the synaptic cleft, and is modeled as $\phi_{ij}(V_j) = 1/\big(1+e^{-\beta_{ij}(V_j-V_{\mathrm{th},ij})}\big)$~\cite{Wicks1996,Kunert2014}. External stimuli in the form of currents $I_{\mathrm{ext},i}$ injected in neurons are added to Eq.~\eqref{eq:kunertv} as $-I_{\mathrm{ext},i}/C_i$, where $C_i$ is the membrane capacitance of neuron $i.$

We obtain expressions for the equilibrium Green's function of the system by linearizing Eqs.~\eqref{eq:kunertv} and~\eqref{eq:kunerts} around the equilibrium of the membrane potentials. With  $\Delta V_j$ and $\Delta s_{ij}$ being the deviations from equilibrium, we obtain $\Delta V_i(t) = \big(g_{0,ij}^\mathrm{g}*\Delta V_j\big)(t) + \big(g_{0,ij}^\mathrm{s}*\Delta s_{ij}\big)(t)$ and  $\Delta s_{ij}(t) = \big(\sigma_{0,ij} * \Delta V_j \big)(t)$. The total direct Green's function ($V_i\leftarrow V_j$) is $g_{0,ij}=g_{0,ij}^\mathrm{g}+g_{0,ij}^\mathrm{s}*\sigma_{0,ij}$, with $\Delta V_i(t) = \big(g_{0,ij} * \Delta V_j)(t)$. (Lower case means direct. See Supplement~\cite{Supplement} for the full expression.)

There are three sources of nonlinearities that can be added back: the saturation of the postsynaptic current when the postsynaptic membrane potential approaches the reversal potential of the ionotropic receptor through which the current flows (Eq.~\eqref{eq:kunertv}); the saturation of the synaptic activity (which can range between $0$ and $1$) due to the finite number of receptors; and the sigmoidal dependence $\phi(V_j)$ of the vescicle release on the presynaptic potential (Eq.~\eqref{eq:kunerts}).

We obtain the full expression for the nonequilibrium $\sigma_{ij}[\bm{V}](t,t')$, the Green's function bringing from $V_j(t)$ to $s_{ij}(t)$, reinserting the nonlinear terms in the equations,
\begin{equation}
\label{eq:sigma_fullyint}
\sigma_{ij}(t,t') = \frac{\sigma_{0,ij}(t-t')}{\partial_{V_j} \phi\big|_{\mathrm{eq}}} \frac{\Delta\phi\big(V_j(t')\big)}{\Delta V_j(t')} \biggl(1-\frac{(\sigma_{ij}*\Delta V_j)(t')}{1-s_{ij,\mathrm{eq}}}\biggr).
\end{equation}
The nonequilibrium $g_{ij}[\bm{V}](t,t')$ is
\begin{equation}
\label{eq:f_fullyint}
g_{ij} (t,t') = g_{0,ij}^\mathrm{g}(t-t') +
g^\mathrm{s}_{0,ij}*\biggl(\bigg(1- \frac{\Delta V_i}{V_{i,\mathrm{eq}}-E_{ij}} \bigg)\sigma_{ij}\biggr)(t,t').
\end{equation}
In the following examples, we will only consider the nonlinear contribution coming from $\sigma_\ab(t,t')$, while we will keep the equilibrium $g_{0,\ab}(t-t')$. The nonequilibrium response function $\chi$ defined by $\delta s_\ab(t) = \chi_\ab(t,t')*\delta V_\beta(t)$, with $\delta V_\beta$ on top of the current state $V_{\beta,\mathrm{eq}} + \Delta V_\beta$, is given by
\begin{equation}
\label{eq:chi}
\begin{split}
\chi_\ab&(t,t') = \frac{\sigma_{0,\ab}(t-t')}{\partial_{V_\beta}\phi\big|_{\mathrm{eq}}} \partial_{V_\beta}\phi\big|_{t'} \bigg(1-\frac{(\sigma_\ab*\Delta V_\beta)(t')}{1-s_{\ab,\mathrm{eq}}}\bigg) \\
&-\int_{t'}^{t} dq \frac{\sigma_{0,\ab}(t-q)}{\partial_{V_\beta}\phi\big|_\mathrm{eq} (1-s_{\ab,\mathrm{eq}})} \Delta \phi (\Delta V_\beta(q)) \chi_\ab(q,t'),
\end{split}
\end{equation}
with the direct response function $f_\ab = g^\mathrm{s}_{0,\ab}*\chi_\ab$.

\paragraph{Illustrative examples}

\begin{figure}[t]
\centering
\includegraphics[width=1\columnwidth]{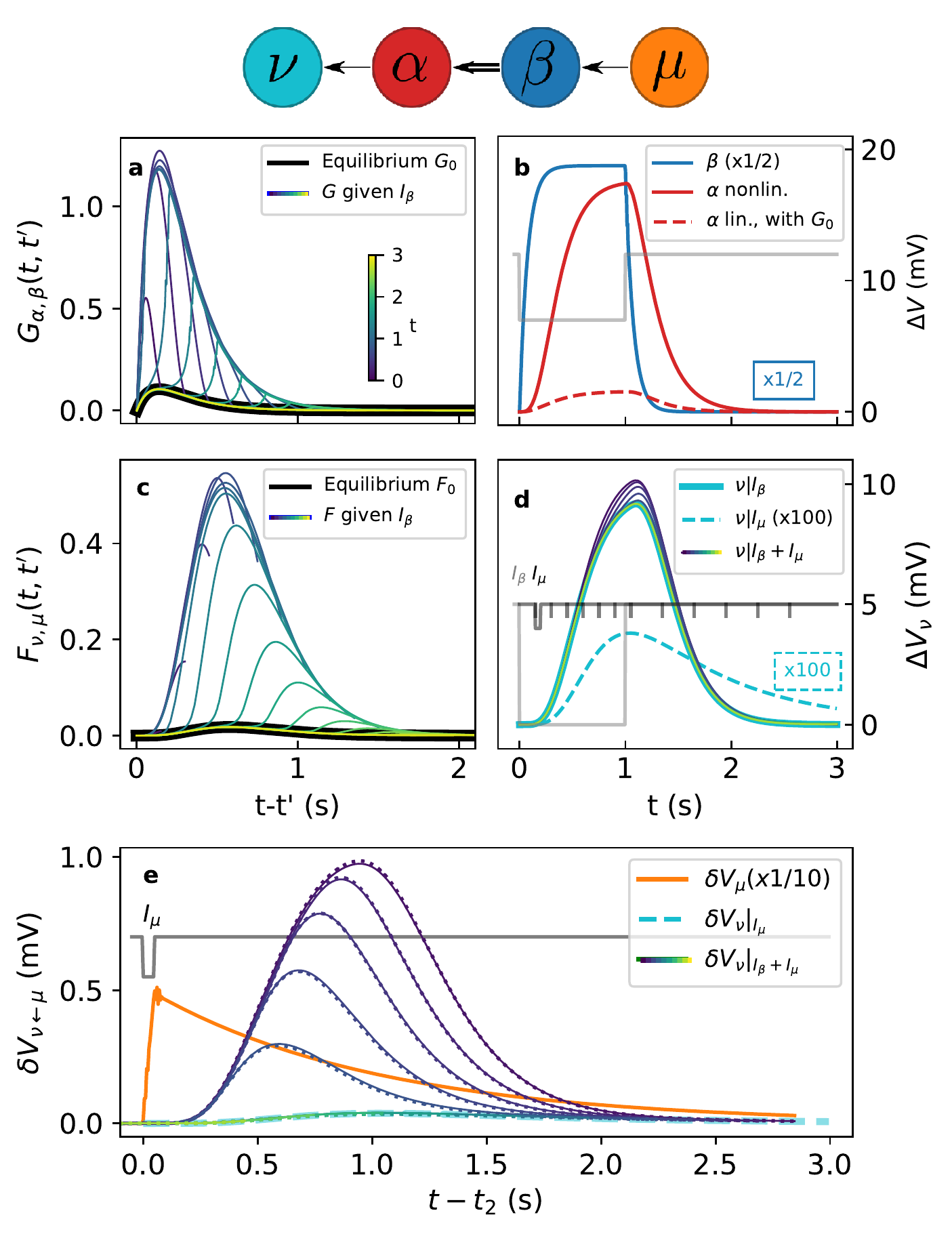}
\caption{
\textbf{top} Scheme of the network.
\textbf{a} Nonequilibrium and equilibrium $G_{(0,)\alpha,\beta}(t,t')$ for selected times $t$ (colored and black curves, respectively).
\textbf{b} $\Delta V_\beta(t)$ (blue), and $\Delta V_\alpha(t)$ obtained with $G_\ab$ (solid red) and with $G_{0,\ab}$ (dashed red). The gray curve is $I_\beta$, with a baseline of $0$ and a peak of $-0.5$ pA (axis not shown).
\textbf{c} Nonequilibrium and equilibrium $F_{(0,)\nu,\mu}(t,t')$ for selected times $t$ (colored and black curves, respectively). Colors as in a.
\textbf{d} $\Delta V_\nu$ obtained with stimulations $I_\beta$ (solid cyan), $I_\mu$ (dashed cyan, x100), and $I_\beta$ + $I_\mu$ (thin lines, colors as in a). Gray curve as in panel b. Black curve: current $I_\mu$, with a baseline of $0$ and a peak of $-0.1$ pA. Black ticks: times $t_2$ at which $I_\mu$ is injected.
\textbf{e} $\delta V_{\nu\leftarrow\mu}$ induced by $I_\mu$ on top of the nonequilibrium state, calculated as $F_{\nu\mu}*\delta V_\mu\big|_{I_\mu}$ (dotted lines) and as $\Delta V_\nu\big|_{I_\beta+I_\mu} - \Delta V_\nu\big|_{I_\beta}$ (solid lines), for different times $t_2$ (colors as in a, $t_2$ as the black ticks in d). $\Delta V_\nu\big|_{I_\mu}$ (cyan dashed line, as in panel d). $\delta V_\mu|_{I_\mu}$ (orange line, x1/10) as produced by perturbation $I_\mu$ (black line) . 
}
\label{fig:gating}
\end{figure}
We provide numerical examples in two simple networks so that results can be understood intuitively. 
In the first example, we show how $F(t,t')$ correctly captures the responses of the neurons to arbitrary stimulations. The example describes a form of gating in a simple feed-forward network with excitatory synaptic connections $\nu\leftarrow\alpha\Leftarrow\beta\leftarrow\mu$, where $\alpha\Leftarrow\beta$ is the only edge where we consider a nonlinearity (as depicted at the top of Fig.~\ref{fig:gating}). We choose parameter values similar to those in Ref.~\cite{Kunert2014} (see Supplement~\cite{Supplement} for more details). The main difference is that $V_{\mathrm{th},\alpha\beta}$ is set to $-10\,\mathrm{mV}$~\cite{Juusola1996} so that the resting potential of neuron $\beta$ sits at the bottom of the sigmoid $\phi_{\alpha\beta}(V_\beta)$. Therefore, small perturbations around the resting potentials of neurons upstream of the nonlinear edge $(\alpha,\beta)$ produce only small responses downstream of that edge, as shown in Figure~\ref{fig:gating}a and c, where the black curves show the equilibrium $G_{0,\alpha\beta}$ and $F_{0,\nu\mu}$ ($=G_{0,\nu\mu}$), respectively.

\begin{figure}[t]
\centering
\includegraphics[width=1\columnwidth]{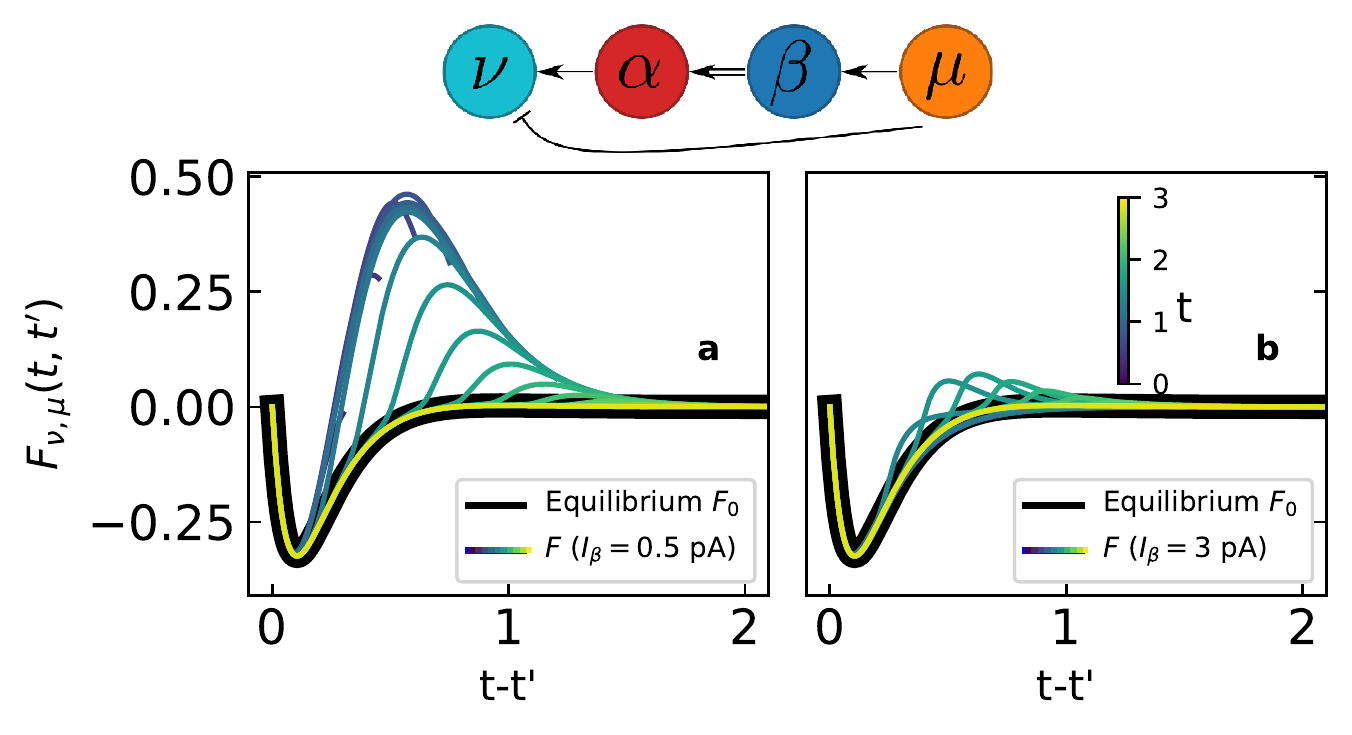}
\caption{Nonequilibrium and equilibrium $F_{(0,)\nu,\mu}(t,t')$ for selected times (colored and black curves, respectively), for currents into $\beta$ of 0.5 pA (\textbf{a}) and 3 pA (\textbf{b}). Scheme of the network (\textbf{top}).}
\label{fig:path_dominance}
\end{figure}

The situation is different if there is a significant change of $V_\beta$, as could happen, for example, due to the application of an odor sensory stimulus. To simulate this, we inject a $0.5\,\mathrm{pA}$ external current $I_\beta$ into $\beta$ for $1$ s (gray curve in Fig.~\ref{fig:gating}b), which induces a $\Delta V_\beta$ as shown in Fig.~\ref{fig:gating}b (blue curve). As a consequence, $\phi_{\alpha,\beta}$ increases significantly and makes $G_{\alpha\beta}$ transiently larger than $G_{0,\alpha\beta}$, as shown in Fig.~\ref{fig:gating}a for selected times $t$. A larger $G_{\alpha\beta}$ allows the activity in $\beta$ to reach $\alpha$ more efficiently (Fig.~\ref{fig:gating}b solid red curve), compared to $G_{0,\alpha\beta}$ (Fig.~\ref{fig:gating}b dashed red curve), and consequently also other neurons downstream of the edge $(\alpha,\beta)$.

In the time interval in which $G_\ab$ is enhanced, any other small perturbations upstream of the nonlinear edge can propagate more effectively to nodes downstream of the edge, compared to at equilibrium. For example, the response function $F_{\mu,\nu}$ from upstream neuron $\mu$ to downstream neuron $\nu$ is shown in in Fig.~\ref{fig:gating}c. 

The nonequilibrium response functions obtained in the simulation via Eqs~\eqref{eq:fullyint} and~\eqref{eq:chi} allow one to compute the response to arbitrary (small) perturbations without solving the underlying differential equations again (see Supplement~\cite{Supplement} for a discussion of how small). 
In contrast, previous approaches required explicitly including the additional perturbations in the main simulation and solving the differential equations. That approach is more computationally expensive and gives less insight because the results depends on the specific perturbation chosen, while our approach gives a characterization for any perturbation. As an illustration, we proceed both ways and compare the results.

To produce a perturbation $\delta V_\mu$ on top of the nonequilibrium state, we consider a shorter current pulse $I_\mu$ of $0.1$ pA ($0.05$ s) injected in neuron $\mu$ (black curve in Fig.~\ref{fig:gating}d) at different times $t_2$ (black ticks). The responses $\Delta V_\nu \big|_{I_\beta+I_\mu}$ produced in neuron $\nu$, explicitly calculated with both $I_\beta$ and $I_\mu$, are shown as the thin curves in Fig.~\ref{fig:gating}d for different $t_2$ (with the same color mapping as panels a and c), together with $\Delta V_\nu\big|_{I_\beta}$ produced by perturbation $I_\beta$ only (solid cyan line). The cyan dashed line, instead, shows the $\delta V_\nu\big|_{I_\mu}$ that the same perturbation would induce with the equilibrium response function $G_{0,\nu\mu}$, multiplied by a factor of $100$.

In Fig.~\ref{fig:gating}e we compare the results obtained with the explicit calculation and the response function $F_{\mu\nu}$, aligning them in time by plotting them vs. $t-t_2$. As a reference, the grey curve shows when $I_\mu$ is applied, and the orange curve the induced $\delta V_\mu$. The solid lines (blue to yellow) are the responses 
$\delta V_{\nu\leftarrow\mu}$ due only to $\delta V_\mu$ and calculated explicitly as $\Delta V_\nu\big|_{I_\beta+I_\mu}-\Delta V_\nu\big|_{I_\beta}$. The dotted lines are instead the same responses $\delta V_{\nu\leftarrow\mu}$ calculated using the response functions as $F_{\nu\mu}*\delta V_\mu|_{I_\mu}$. The two calculations show close agreement. 

The gating effect is clear in this plot: As $I_\mu$ ceases to be in coincidence with $I_\beta$, its enhanced effect becomes smaller and finally vanishes when $I_\mu$ is applied after $V_\beta$ is back to the resting value. This is also represented by the response functions in Fig.~\ref{fig:gating}c. To draw a parallel with the SLDS~\cite{Linderman2019,Costa2019}, the state induced by the perturbation corresponds to the switching from the equilibrium linear system to another linear system with different parameters.

A second example calculation illustrates how effective interactions can also change dramatically, e.g. from an inhibitory connection to a connection that computes a fractional derivative of $\delta V_\mu(t)$. We modify the network used above by adding an inhibitory synapse $\nu\vdash\mu$, so that there are two paths from $\mu$ to $\nu$, a direct inhibitory path and an indirect excitatory one that goes through $\alpha\Leftarrow\beta$ (for the parameters, see the Supplement~\cite{Supplement}).

At equilibrium, the effective response function $F_{0,\nu\mu}(t-t')$ ($=G_{0,\nu\mu}$) is purely inhibiting (black curve in Fig.~\ref{fig:path_dominance}a,b), because $G_{0,\alpha\beta}$ is very small (as in the previous example). When the system is perturbed by the same square current pulse $I_\beta$ flowing into neuron $\beta$ as above, the Green's function of the edge $\alpha\Leftarrow\beta$  is enhanced, and as a consequence $F_{\nu\mu}(t,t')$ transiently acquires the shape of a fractional derivative-like kernel shown in Fig.~\ref{fig:path_dominance}a, before decaying back to the equilibrium $F_{0,\nu\mu}$.

This effect disappears if $\beta$ is stimulated too strongly, as shown in Fig.~\ref{fig:path_dominance}b for a current of 3 pA. As the $(\alpha,\beta)$ synapse reaches the top of $\phi$ and saturates, it becomes again unable to transmit additional perturbations. The analysis reveals how $\beta$'s activity influences signal propagation from $\mu$ to $\nu$ in a non-trivial way. Such a computation might exist in the brain to  integrate different sensory stimuli.  In our odor stimulus analogy,   activation of sensory neuron $\beta$ by odorant B would adjust functional connectivity to modulate the animal's downstream response to a second stimulus M in $\mu$.  Low  or high concentrations of odor B would have no effect, but intermediate concentrations would cause the animal to respond to the derivative of odor M.

In conclusion, we have presented an equation for nonequilibrium Green's functions to describe time-dependent and nonlinear networks of neurons. We believe this approach will prove very useful for two reasons. First, it provides a bridge between biophysical-like models of neural networks and their effective counterparts. Second, it allows one to isolate and understand the role of specific sets of neurons in modulating the functional connectivity of neural networks, especially in contexts like \textit{C. elegans} in which the most significant nonlinearities may be localized in specific degrees of freedom or edges. We have illustrated these concepts with two numerical examples that show how a nonlinear edge can modify in a time-dependent way the interaction between other neurons, both quantitatively and qualitatively. We ran the calculations for these examples on very simple networks. But, since the calculations deal with the time-evolution of the effective ``connected'' Green's function, they hold whether the paths are direct, indirect, or involve recurrence. Therefore, the illustrated examples are representative of the effect that nonlinearities associated with hub neurons can have on large portions of the functional connectivity.

\paragraph*{Acknowledgements} We thank Martin Eckstein and Fulvio Parmigiani for the insightful discussions, and Carlos Brody, Kevin S. Chen, and Ross Dempsey for the critical reading of the manuscript. F.R. was supported by the Swartz Foundation via the Swartz Fellowship for Theoretical Neuroscience. This work was supported in part by the National Science Foundation, through the Center for the Physics of Biological Function (PHY-1734030), and by the National Institute of Neurological Disorders and Stroke of the National Institutes of Health under New Innovator Award number DP2NS116768 to A.M.L. The content is solely the responsibility of the authors and does not necessarily represent the official views of the National Institutes of Health.


\begin{thebibliography}{100}
\bibitem{footnotem1}
And if $j$ is the only neuron setting the boundary conditions, in practice, if it is the only one being externally perturbed.
\end{thebibliography}
\end{document}


\title{Supplemental material for: Nonequilibrium Green's functions for functional connectivity in the brain}

\begin{abstract}
A theoretical framework describing the set of interactions between neurons in the brain, or functional connectivity, should include dynamical functions representing the propagation of signal from one neuron to another. Green's functions and response functions are natural candidates for this but, while they are conceptually very useful, they are usually defined only for linear time-translationally invariant systems. The brain, instead, behaves nonlinearly and in a time-dependent way. Here, we use nonequilibrium Green's functions to describe the time-dependent functional connectivity of a continuous-variable network of neurons. We show how the connectivity is related to the measurable response functions, and provide two illustrative examples via numerical calculations, inspired from \textit{C. elegans}.
\end{abstract}
\author{Francesco Randi}
\affiliation{Department of Physics, Princeton University, Jadwin Hall, Princeton, NJ 08544, USA}
\author{Andrew M. Leifer}
\affiliation{Department of Physics, Princeton University, Jadwin Hall, Princeton, NJ 08544, USA}
\affiliation{Princeton Neuroscience Institute, Princeton University, NJ 08544, USA}
\date{\today}
\maketitle

\section{Notation}
$g$ is for Green's functions, $f$ is for response functions. Lower case is for the direct paths, upper case for the effective ones. Subscript 0 stands for time-translationally invariant. The superscript on $G_0$ denotes the indices that are excluded from the summation in the definition of $G_0$.

Green's functions:
\begin{itemize}
\item $g_{0,ij}(t-t')$: equilibrium direct Green's function for the pair of neurons $i\leftarrow j$
\item $G_{0,ij}(t-t')$: equilibrium effective Green's function for neurons $i\leftarrow j$ in a network
\item $g_{ij}(t,t')=g_{0,ij}(t-t')+\pi_{ij}[\bm{\psi}](t,t')$: nonequilibrium direct Green's function written as the sum of the equilibrium $g_{0,ij}$ and a nonequilibrium term $\pi_{ij}$ that depends on the state $\bm{\psi}$ of the system. 
\item $G_{ij}(t,t')$: nonequilibrium effective Green's functions for neurons in a network
\end{itemize}

Response functions
\begin{itemize}
\item $f_{0,ij}(t-t')$: equilibrium direct response function for an isolated pair of neurons (equal to $g_{0,ij}$)
\item $F_{0,ij}(t-t')$: equilibrium effective response function for two neurons in a network (equal to $G_{0,ij}$)
\item $f_{ij}(t,t')=f_{0,ij}(t-t') + \overline\chi_{ij}[\bm{\psi}](t,t')$: nonequilibrium direct response function for an isolated pair of neurons, written as the sum of the equilibrium response function $f_{0,ij}$ and a nonequilibrium term $\overline\chi_{ij}$ which depends on the state $\bm{\psi}$ of the system.
\item $F_{ij}(t,t')$: nonequilibrium response function for neurons in a network
\end{itemize}

\section{Linear and time-translationally invariant case}
\subsection{Isolated pair of neurons}
In the linear and time-translationally invariant (TTI) case, the activity $\psi_i(t) = \psi_{i,\mathrm{eq}}+\Delta\psi_i(t)$ of neuron $i$ is given by
\begin{equation}
\label{eq:lin_time_inv}
\psi_i(t) = \psi_{i,eq} + \sum_{j\in \text{ all}} g_{0,ij}*\Delta\psi_j\, (t) + g_{0,i}^\mathrm{ext}*I^\mathrm{ext}_i(t),
\end{equation}
where the $\psi_\mathrm{eq}$ are the resting activity levels of the neurons (which depend on the equilibrium inputs from the other neurons) and the $\Delta \psi$ are the changes of the activity with respect to those resting values. $*$ denotes a convolution such that
\begin{equation}
\begin{split}
\big(g*\Delta\psi\big)(t) &= \int dt_1\,g(t,t_1) \Delta\psi(t_1) \\
g(t,t') &= g(t-t') \text{ for TTI case}
\end{split}
\end{equation}
$g_{0,ij}(t-t')$ is the Green's function, or transfer function, describing the direct interaction $i\leftarrow j$ from neuron $j$ to neuron $i$ (the subscript $0$ denotes their time-translational invariance). $I_i^\mathrm{ext}$ is an external input to $j$, and $g_{0,i}^\mathrm{ext}$ is the transfer function needed to calculate the activity induced by $I_i^\mathrm{ext}$. Note that we use $\Delta\psi$ to denote the deviation of the current state from the equilibrium activity, and $\delta\psi$ to denote perturbations induced in order to probe the properties of the system (see below).

If we consider an isolated pair of neurons $i$ and $j$, not connected to any network and with a one-way connection $i\leftarrow j$, then the right-hand side of Eq.~\eqref{eq:lin_time_inv} for neuron $i$ contains a convolution only for the single neuron $j$. If we know $\Delta \psi_j$, and $\Delta\psi_j$ entirely sets the boundary conditions of the problem because, e.g. it is the only neuron being perturbed, the $\psi_i(t)$ calculated from Eq.~\eqref{eq:lin_time_inv} is exact. In this situation, $j$ would not have any input but $I_j^\mathrm{ext}$, and would be $\Delta\psi_j(t) = \psi_{j,\mathrm{eq}} +g_{0,j}^\mathrm{ext}*I^\mathrm{ext}_j(t)$

A strategy to probe the dynamical properties of the two-neurons system in a given state $\bm{\psi} = \bm{\psi_\mathrm{eq}}+\bm{\Delta\psi}$ is to measure its response functions. $\bm{\Delta\psi}$, here, has been produced in an unspecified way: for example by an external perturbation, a sensory stimulus, or a change in the cellular properties that has determined a change in the resting activity of the neuron with respect to the usual equilibrium. To measure the response function, we can induce in $j$ an additional perturbation $\delta \psi_j$ via an external current $I_j^\mathrm{ext}$ and measure the response that this produces in $i$. This response can be expressed as
\begin{equation}
\label{eq:lin_time_inv_f}
\delta\psi_i(t) = f_{0,ij}*\delta\psi_j\,(t).
\end{equation}
$f_{0,ij}(t-t')$ is the response function of the edge $(i,j)$, and in the linear and time-translationally invariant case is equal to the Green's function $g_{0,ij}(t-t')$. Moreover, since the system is linear, Eq.~\eqref{eq:lin_time_inv_f} is valid for $\delta\psi$s of any magnitude.

\subsection{Neurons in a network}
Eq.~\eqref{eq:lin_time_inv} only describes direct connections between the neurons. If the pair $(i,j)$ is embedded in a network, the signal can travel from $j$ to $i$ both on direct and indirect paths. Therefore, if the boundary conditions are set by $\Delta\psi_j$ as above, in this case Eq.~\eqref{eq:lin_time_inv} yields a result that is correct only up to the first perturbative order. To obtain the exact result, Eq.~\eqref{eq:lin_time_inv} should be solved for each neuron in the network, and the values of $\psi(t)$ updated at each time-step $t$ for each neuron.

However, the direct Green's functions $g_{0,ij}$ of a linear system can be condensed in effective, or connected, Green's functions $G^j_{0,ij}(t-t')$, which in the context of Volterra integral equations are the so-called resolvent kernels~\cite{Linz1985} (for the meaning of the superscript $j$ see below). If we know $\Delta\psi_j$ and if we want to calculate the activity $\Delta\psi_i$ produced in $i$ without calculating $\Delta\psi$ for the whole network (and if the boundary conditions are set by $\Delta\psi_j$), we can recursively insert the contribution from all the neurons (and therefore all the paths) in Eq.~\eqref{eq:lin_time_inv} to obtain such connected Green's function. The goal is to write
\begin{equation}
\label{eq:lin_time_inv_deltapsi}
\psi_i(t) = \psi_{i,\mathrm{eq}} + G^j_{0,ij}*\Delta\psi_j(t).
\end{equation}
Focusing on the $\Delta\psi_i$ in Eq.~\eqref{eq:lin_time_inv}, dropping for the moment the external inputs, we have
\begin{equation}
\Delta\psi_i(t) = g_{0,ij}*\Delta\psi_j + \sum_{\mu\neq j} g_{0,i\mu}*\Delta\psi_\mu.
\end{equation}
The $\Delta\psi_\mu$s are themselves only determined by $\Delta\psi_j$, so that we can substitute $\Delta\psi_\mu$ with $g_{0,\mu j}*\Delta\psi_j+\sum_\nu g_{0,\mu\nu}*\Delta\psi_\nu$
\begin{equation}
\begin{split}
\Delta\psi_i(t) &= g_{0,ij}*\Delta\psi_j + \sum_{\mu\neq j} g_{0,i\mu}*g_{0,\mu j}*\Delta\psi_j\\
&+\sum_{\mu\neq j}\sum_{\nu} g_{0,i\mu}*g_{0,\mu\nu}*\Delta\psi_\nu.
\end{split}
\end{equation}
Recursively repeating this step, we obtain
\begin{equation}
\begin{split}
\Delta\psi_i(t) &= \biggl\{ g_{0,ij} +\sum_{\mu\neq j} g_{i\mu}*\\
&\biggl[g_{0,\mu j}+\sum_{\nu\neq j} g_{0,\mu\nu}*g_{0,\nu j} + ...
\biggr]\biggr\}*\Delta\psi_j\,(t).
\end{split}
\end{equation}
$G^j_{0,ij}$ can be read off the above equation comparing it with Eq.~\eqref{eq:lin_time_inv_deltapsi}. The superscript $j$ denotes the fact that $j$ is excluded in the sums. 

Additionally, noting that the terms inside the square brackets are $G^j_{0,\mu j}$, we obtain an equation relating the direct and connected Green's functions:
\begin{equation}
\label{eq:fullyconn}
G^j_{0,ij} = g_{0,ij} + \sum_{\mu\neq j} g_{0,i\mu} * G^j_{0,\mu j}.
\end{equation}

In a linear system, one would write the above equation in Fourier space, where the convolutions are transformed in simple products, like in Ref.~\cite{Brinkman2018}. But this cannot be done when the functions are not time-translationally invariant, as is the case in the main discussion of this work.

Because we already take into account indirect paths via $G_{0}$, in Eq.~\eqref{eq:lin_time_inv_deltapsi} the convolution is summed only over the neurons that set the boundary conditions. In the simple example in which an external current is injected in a single neuron $j$, the convolution should be performed only with $\Delta\psi_j$. If an external current is injected in two neurons, the convolutions in Eq.~\eqref{eq:lin_time_inv_deltapsi} should be with the activities produced by those external currents separately.

As in the previous section, we can characterize the system measuring its response functions. This can be done experimentally by inducing a perturbation $\delta\psi_j(t)$ and measuring the produced $\delta\psi_i(t)$. Also for neurons in a network, in the linear and time-translationally invariant case the response functions $F_{0,ij}$ are equal to the Green's functions $G_{0,ij}^j$, such that one can write
\begin{equation}
\delta\psi_i(t) = F_{0,ij}*\delta\psi_j\,(t)
\end{equation}

The activity produced by an external current $I_j^\mathrm{ext}$ in $j$ also contains a network term if the neurons are in a network. With a recursive procedure similar to the one leading to Eq.~\eqref{eq:fullyconn}, one obtains that the $\Delta\psi_j$ induced by $I_j^\mathrm{ext}$ is 
\begin{equation}
\label{eq:lin_time_inv_G_ext}
\Delta\psi_j(t) = (\delta(t,t')+G_{0,jj})*g_{0,j}^\mathrm{ext}*I_j^\mathrm{ext}.
\end{equation}
$G_{0,jj}$ does not have a superscript $j$, which means that it should be calculated as
\begin{equation}
G_{0,jj} = g_{0,jj} + \sum_{\mu\in\text{ all}} g_{0,j\mu} * G_{0,\mu j}.
\end{equation}
The next section explains why in some cases some indices have to be excluded from the sums and in others they don't.

\subsection{Self loops, \textit{measured} activities, and alternative formulation}
The goal of writing $\Delta\psi_i = G_{0,ij}^j*\Delta\psi_j$, and $\delta\psi_i = F_{0,ij} * \delta\psi_j$, is to have expressions that relate \emph{measured} activities. If we know $\Delta\psi_j(t)$ because we measured it, $\Delta\psi_j$ already contains the all the  corrections due to the presence of the network. In the summation in Eq.~\eqref{eq:fullyconn}, $\mu$ cannot be equal to $j$ because otherwise we would be considering paths that start from $j$ and return to $j$ before coming to $i$. These paths affect $\Delta\psi_j$ because they bring input to $j$, so their contribution has already been accounted for in the known $\Delta\psi_j$. Excluding $j$ from the summation in Eq.~\eqref{eq:fullyconn} is equivalent to considering paths that start from $j$ and end on $i$ without ever coming back to $j$.

When we want to calculate, instead, the activity $\Delta\psi_j$ produced by an external current injected into $j$ (Eq.~\eqref{eq:lin_time_inv_G_ext}), we need to consider those network effects on $j$ itself that we discarded above. This is why in Eq.~\eqref{eq:lin_time_inv_G_ext}, $G_{0,ij}$ appears without a superscript $j$.

\paragraph{Alternative formulation} An alternative formulation in which no index is excluded from all $G_{0,ij}$s is possible, but comes with a crucial disadvantage. Let us call $\Delta\overline\psi_j (t)$ the activity that an external stimulation $I^\mathrm{ext}_j$ would produce in neuron $j$ if it were \emph{isolated}
\begin{equation}
\Delta\overline\psi_j(t) = g_{0,j}^\mathrm{ext}*I_{j}^\mathrm{ext}.
\end{equation}
The activity of that same neuron $j$ considering the feedback from the network, would be given by $\Delta\psi_j = (\delta(t,t') + G_{0,jj})*\Delta\overline\psi_j$ (the same as Eq.~\eqref{eq:lin_time_inv_G_ext}). The activity of other neurons $i$ due to $\Delta\psi_j$ would now be given by $\Delta\psi_i(t)=G_{0,ij}*\Delta\overline\psi_j(t)$, with the summation in $G_{0,ij}$ running over all indices.

However, the relation now is between the measured $\Delta\psi_i$ and $\Delta\overline\psi_j$, the activity of neuron $j$ if it were isolated from the network, which cannot be measured. To make use of this relation to obtain information about the network, we need to make assumptions about $\Delta\overline\psi_j$, which translate into assumptions about $g_{0,j}^\mathrm{ext}$ and $I_j^\mathrm{ext}$. In some cases, $g^\mathrm{ext}$ and $I^\mathrm{ext}$ can be estimated or separately characterized, like in the case of optogenetic stimulations with light-gated ion channels, or channelrhodopsins ($I^\mathrm{ext}$ depends, for example, on the gating kinetics of the channelrhodopsin). Given that the stimulus is an electrical current, the coupling to the membrane potential of a neuron is via a first-order equation and $g^\mathrm{ext}\propto e^{-t/\tau}$. 

This characterization is not general, however, because in the same nervous system stimuli might be applied to the neurons also via other types of molecules, like G-protein coupled receptors. With these receptors, the coupling to the neuron's activity is likely to be at least via a second-order equation because there are intermediate steps involved, so that $g^\mathrm{ext}$ is not a simple exponential anymore.

Depending on the context and the goal of the calculations, be they for numerical simulations or data analysis, either of the two formulations (with $\Delta\overline\psi$ or $\Delta\psi$) can be used. In the following, we will keep working with the \emph{measured} $\Delta\psi$.

\section{Nonequilibrium case}
\subsection{Isolated pair of neurons}
Also in the nonlinear case we would like to formally write the state of the system as a convolution of a Green's function and the past activities of the neurons:
\begin{equation}
\label{eq:neq_direct}
\Delta\psi_i(t) = \int dt'\,g_{ij}(t,t')\Delta\psi_j(t') = g_{ij}*\Delta\psi_j\,(t).
\end{equation}
In contrast to the linear and time-translationally invariant case, the Green's function $g_{ij}$ in Eq.~\eqref{eq:neq_direct} is not a time-invariant kernel (hence, nonequilibrium). Instead it is, in principle, functionally dependent on the whole state of the system $\bm{\psi}$. We can write $g_{ij}[\bm{\psi}]$ as the sum of a linear and time-translationally invariant term $g_{0,ij}(t-t')$ and a nonequilibrium term $\pi_{ij}(t,t')$:
\begin{equation}
\label{eq:neq_direct_gf}
g_{ij}[\bm{\psi}](t,t') = g_{0,ij}(t-t') + \pi_{ij}[\bm{\psi}](t,t').
\end{equation}

There is no general recipe for calculating $\pi_{ij}[\bm{\psi}]$. Different models describing the nonlinear or time-dependent nature of the edge $(i,j)$ will involve their own specific solutions. However, Eq.~\eqref{eq:neq_direct} and~\eqref{eq:neq_direct_gf} are useful as formal devices because they will allow us to express how a given $\pi_{ij}[\bm{\psi}]$, once it is known, should be used to calculate the modified properties of the rest of the network. 

\subsection{Response function of isolated pair}
\label{sec:neq_f}
In the nonequilibrium case, the response function $f_{ij}$ is, in general, not equal to the Green's function $g_{ij}$. If the system is in the state $\bm{\psi} = \bm{\psi_\mathrm{eq}}+\bm{\Delta\psi}$ (where $\bm{\Delta\psi}$ has been produced in an unspecified way), and we induce a perturbation $\delta\psi_\beta$ in $\beta$ to probe the system, the activity $\delta\psi_\alpha$ produced in $\alpha$ can still be written as 
\begin{equation}
\label{eq:neq_f_psi}
\delta\psi_\alpha(t) = f_{\ab}*\psi_\beta\,(t),
\end{equation}
but here $f_{\ab}$ is given by
\begin{equation}
\label{eq:nonlinedge}
f_\ab[\bm{\psi}](t,t') = \frac{\delta\psi_\alpha}{\delta\psi_\beta} = g_{0,\ab}(t-t') + \overline\chi_\ab[\bm{\psi}](t,t'),
\end{equation}
where $\overline\chi_\ab[\bm{\psi}]$ is the nonequilibrium part of the response function and is given by
\begin{equation}
\label{eq:chibar}
\overline\chi_\ab[\bm{\psi}](t,t') = \pi_\ab(t,t') + \biggl(\frac{\delta\pi(t,t_1)}{\delta\psi_\beta(t')}*\psi_\beta(t_1)\biggr)(t,t').
\end{equation}

Also here, $f_{ij}$ is not a time-translationally invariant transfer function, and depends on the state of the system $\bm{\psi}$ via $\overline\chi_\ab[\bm{\psi}]$. Therefore, it has to be calculated according to its full expressions (or approximations to it). But, provided that $\delta\psi_j$ is ``small'' (see below), once $\overline\chi_\ab$ and $f_{ij}$ are computed, Eq.~\eqref{eq:neq_f_psi} can be used to to compute the response $\delta\psi_\alpha$ to any perturbation $\delta\psi_\beta$.

In neurons, we can expect the linear-response regime to be valid also for $\delta\psi$ of significant magnitude. In the most general nonlinear system, linear responses around an arbitrary state can be very limited approximations. But nonlinearities like sigmoidal thresholds at the synapses have large activity intervals where they are well-approximated by a linear function. In the examples in the main text, at equilibrium the synapse $(\alpha,\beta)$ ``sits'' at the bottom of $\phi(V_\beta)$, the sigmoidal function describing the dependence of the synaptic vesicle release on the calcium influx into the presynaptic site. As the system is brought out of equilibrium, the synapse reaches the central region of the sigmoid $\phi(V_\beta)$, where the sigmoid is again almost linear and where the linear-response approximation works well.

Two cases can be discussed starting from Eq.~\eqref{eq:chibar}. $\pi(t,t')$ can be nonequilibrium in two ways: it can be time-dependent or explicitly nonlinear (or a combination of the two). If $\pi$ is purely time-dependent and does not depend on $\psi_\beta$, the only nonzero term in the right-hand side of Eq.~\eqref{eq:chibar} is $\pi$. In this case, the response function is equal to the Green's function: $f_\ab = g_{0,\ab} + \pi_\ab$. If $\pi$ depends on $\delta\psi_\beta$, then also the second term is nonzero and the $f_\ab\neq g_\ab$.

\subsection{Neurons in a network: nonequilibrium Green's functions}
The connected Green's functions for neurons in a network will be determined also by the nonequilibrium terms. To derive an equation for the nonequilibrium Green's functions $G_{ij}$ we start from the following experimental consideration related to \emph{C. elegans} that can be taken as a general starting point.
Characterizations of some synpases in the worm have shown that they are linear throughout a large part of the physiological range of membrane potentials~\cite{Liu2009,Lindsay2011,Narayan2011}. However, we know that nonlinearities and time-dependences are critically important in the \emph{C. elegans} nervous system and in nervous systems generally, because they allow the network to perform  computations, including for example responding  to sensory stimuli in a context dependent manner~\cite{Mochi2019,Dobosiewicz2019}. How does a network have many linear edges but also show widespread nonlinear behaviors? In the integral formulation with nonequilibrium Green's functions it is straightforward to show how these two observations can coexist.

We, therefore, consider a network in which only one of the edges, $(\alpha,\beta)$, displays a significant nonlinearity. This is in contrast to an approach in \cite{Ocker2017} which assumes nonlinearities that are homogeneous over the network and then proceeds with their systematic expansion.

The connected nonequilibrium Green's function for the effective edge $(i,j)$ can be calculated with a recursive procedure similar to the one leading from Eq.~\eqref{eq:lin_time_inv} to Eqs.~\eqref{eq:lin_time_inv_deltapsi} and~\eqref{eq:fullyconn}, but using the nonequilibrium Green's function in Eq.~\eqref{eq:neq_direct_gf} for the edge $(\alpha,\beta)$. The result for $i\neq\alpha,j\neq\beta$ is
\begin{equation}
\label{eq:neq_fullyconn_G}
G_{ij}(t,t') = G_{0,ij}(t-t') + G^j_{0,i\alpha}*\pi_{\ab}[\bm{\psi}]*G_{\beta j}\,(t,t'),
\end{equation}
where $(A*B)(t,t')=\int dt_1\,A(t,t_1)B(t_1,t)$ and the activity $\Delta\psi_i$ can be written as
\begin{equation}
\label{eq:neq_fullyconn_psi}
\Delta\psi_i(t) = \sum_{\text{b.c.}}G_{ij}*\Delta\psi_j,
\end{equation}
with, as usual, a summation on the $\Delta\psi$s that set the boundary conditions (b.c.). Differently from Eq.~\eqref{eq:fullyconn}, here $G_{ij}$ is not a time-translationally invariant transfer function. The nonequilibrium $\pi_\ab[\bm{\psi}]$ and $G_{ij}$ have to be calculated for each time-step taking the nonlinear and time-dependent nature of $\pi_\ab[\bm{\psi}]$ into account, and the calculation depends, therefore, on the specific state $\bm{\psi}$ considered. Once it is calculated, however, it can be plugged in in the expression for all the nonequilibrium $G_ij$ (Eq.~\eqref{eq:neq_fullyconn_G}), and can be used to calculate $\Delta\psi_i$ via convolutions only, without running a nonlinear calculation.

The full nonlinear calculation of $\pi_\ab[\bm{\psi}]$ does not involve, however, all the neurons of the network, in common situations. $\pi_\ab[\bm{\psi}]$ will depend likely on a limited subsets of neurons only. A nonlinearity determined by threshold or saturation at a synapse will depend only on the presynaptic activity $\psi_\beta$. If neuromodulation is involved, it will be determined by a defined set of neurons. 

Additionally, in the calculation of $G_{ij}$ the effect of the network recurrence is condensed in only single additional effective edge $(\beta,j)$. To calculate $G_{ij}$ one needs, therefore, to simultaneously calculate only $G_{\beta,j}$.

There are some special cases for Eq.~\eqref{eq:neq_fullyconn_G}. 

If $(i,j)=(\alpha,\beta)$
\begin{equation}
G_{\ab} = G^\beta_{0,\ab} + (\delta(t,t')+G^\beta_{0,\alpha\alpha})*\pi_{\ab}.
\end{equation}
If $i=\alpha$ and $j\neq\alpha,\beta$
\begin{equation}
G_{\alpha j} = G^j_{0,\alpha j} + (\delta(t,t')+G^j_{0,\alpha\alpha})*\pi_{\ab}*G_{\beta j}.
\end{equation}
If $i,j=\alpha$
\begin{equation}
G_{\alpha\alpha} = G_{0,\alpha\alpha} + (\delta(t,t')+G_{0,\alpha\alpha})*\pi_{\ab}*G_{\beta\alpha}.
\end{equation}
If $i\neq\alpha,\beta$ and $j=\beta$
\begin{equation}
G_{i\beta} = G^\beta_{0,i\beta} + G^\beta_{0,i\alpha}*\pi_{\ab}.
\end{equation}
If $i,j=\beta$
\begin{equation}
G_{\beta\beta} = G_{0\beta\beta} + G_{0,\beta\alpha}*\pi_\ab*(\delta(t,t') + G_{\beta\beta}).
\end{equation}
If $i\neq\alpha$ and $j=\alpha$
\begin{equation}
G_{i\alpha} = G^\alpha_{0,i\alpha}.
\end{equation}
If $i=j$ and $i\neq\alpha,\beta$
\begin{equation}
G_{ii} = G_{0,ii} + G_{0,i\alpha}*\pi_\ab*G_{\beta i}.
\end{equation}

\subsection{Neurons in a network: nonequilibrium response functions}
The response function of an effective edge $(i,j)$ is modified by the network in a similar way as the nonequilibrium $G_{ij}$ in Eq.~\eqref{eq:neq_fullyconn_G}. We assume again one single nonequilibrium edge $(\alpha,\beta)$ and insert recursively the contributions from all the neurons in Eq.~\eqref{eq:lin_time_inv_f}, using Eq.~\eqref{eq:nonlinedge} for the edge $(\alpha,\beta)$. The result for $i\neq\alpha,j\neq\beta$ is
\begin{equation}
\label{eq:fullyint}
F_{ij}(t,t') = F_{0,ij}(t-t') + \big(F_{0,i\alpha}* \overline\chi_\ab* F_{\beta j}\big)(t,t').
\end{equation}
The response $\delta\psi_i$ in $i$ produced by a small perturbation $\delta\psi_j$ in $j$ can be written as
\begin{equation}
\label{eq:neq_fullyint_f_psi}
\delta\psi_i(t) = \int dt'\, F_{ij}(t,t') \delta\psi_j(t').
\end{equation}
Also here $F_{ij}$ is not a time-translationally invariant transfer function, and depends on the state of the system $\bm{\psi}$ via $\overline\chi_\ab[\bm{\psi}]$. But, provided that $\delta\psi_j$ is ``small'' (see discussion in the section about the isolated-pair response function), once $\overline\chi_\ab$ and $F_{ij}$ are computed in the full nonlinear calculation, Eq.~\eqref{eq:neq_fullyint_f_psi} can be used to to compute the response $\delta\psi_i$ to any perturbation $\delta\psi_j$.

Also for the response functions $F_{ij}$, as for the Green's functions $G_{ij}$, there are some special cases of Eq.~\eqref{eq:fullyint}. As a reminder, $F_0=G_0$.

If $(i,j)=(\alpha,\beta)$
\begin{equation}
F_{\ab} = F^\beta_{0,\ab} + (\delta(t,t')+F^\beta_{0,\alpha\alpha})*\overline\chi_{\ab}.
\end{equation}
If $i=\alpha$ and $j\neq\alpha,\beta$
\begin{equation}
F_{\alpha j} = F^j_{0,\alpha j} + (\delta(t,t')+F^j_{0,\alpha\alpha})*\overline\chi_{\ab}*F_{\beta j}.
\end{equation}
If $i,j=\alpha$
\begin{equation}
F_{\alpha j} = F_{0,\alpha\alpha} + (\delta(t,t')+F_{0,\alpha\alpha})*\overline\chi_{\ab}*F_{\beta\alpha}.
\end{equation}
If $i\neq\alpha,\beta$ and $j=\beta$
\begin{equation}
F_{i\beta} = F^\beta_{0,i\beta} + F^\beta_{0,i\alpha}*\overline\chi_{\ab}.
\end{equation}
If $i,j=\beta$
\begin{equation}
F_{\beta\beta} = F_{0\beta\beta} + F_{0,\beta\alpha}*\overline\chi_\ab*(\delta(t,t') + F_{\beta\beta}).
\end{equation}
If $i\neq\alpha$ and $j=\alpha$
\begin{equation}
F_{i\alpha} = F^\alpha_{0,i\alpha}.
\end{equation}
If $i=j$ and $i\neq\alpha,\beta$
\begin{equation}
F_{ii} = F_{0,ii} + F_{0,i\alpha}*\overline\chi_\ab*F_{\beta i}.
\end{equation}

\subsection{Experimental characterization}
$F_{ij}$ are the response functions that can be obtained in experiments on networks of neuron. Importantly, they are always well defined, for two reasons. First, regardless of how nonlinear the system is, an experiment can always be designed to probe the system by perturbing it and measuring its response. Second, their role is always clear even if we are not considering any other neuron in addition to $i$ and $j$, both in an experiment and in theoretical calculations, even if in the latter the response function would be difficult to calculate without making assumptions on the rest of the network.

In an experiment, $F_{ij}$ can be measured as responses to impulsive perturbations. One could object that, by applying an impulsive external stimulation $I^\mathrm{ext}_j$ to neuron $j$, the $\delta\psi_j$ induced in $j$ would be given by Eq.~\eqref{eq:lin_time_inv_G_ext} and would not be itself a delta function. Therefore, the response of neuron $i$ would be given not by $F_{ij}$ alone, but by the convolution of $(1+G_{0,jj})*g_{0,j}^\mathrm{ext}$ with $F_{ij}$. However, if the coupling between the external stimulation $I^\mathrm{ext}_j$ and $\psi_j$ is via a first-order differential equation, e.g. for current in an equation for membrane potential, for which $g_{0,j}^\mathrm{ext}\propto e^{-(t-t')/\tau}$, this limit is of no concern. The timescale $\tau$ is a property of neuron $j$ and applies to any input to that neuron, so that there are no response functions with a path through $j$ that are ``faster'' than $g^\mathrm{ext}_j$.

\section{\textit{C. elegans} examples}
Defining 
\begin{equation}
\bar a_{ij}\equiv a_{d,ij} + a_{r,ij}\phi_{ij}(V_{j,\eq})
\end{equation}
and 
\begin{equation}
\bar \gamma_{i}\equiv \gamma_{i} + \sum_j \gamma_{ij}^\mathrm{g} + \sum_j \gamma_{ij}^\mathrm{s} s_{ij,\eq}
\end{equation}
the equilibrium Green's functions $\sigma_{0,ij}$, $g_{0,ij}^\mathrm{g}$, and $g_{0,ij}^\mathrm{s}$ are
\begin{align}
\sigma_{0,ij}(t) &= \theta(t) a_r (1-s_{ij,\mathrm{eq}}) (\partial_{V_j}\phi_{ij}\big|_\eq) e^{-\bar a_{ij} t},\\
g_{0,ij}^\mathrm{g}(t) &= \theta(t) \gamma_{ij}^\mathrm{g} e^{-\bar \gamma_i t},\\
g_{0,ij}^\mathrm{s}(t) &= \theta(t) \gamma_{ij}^\mathrm{s} (V_{i,\eq} - E_{ij}) e^{-\bar \gamma_i t}.
\end{align}
\paragraph{Numerical calculations}
As parameters for the numerical calculations, we choose $C_i=1$ pF, $E_{c,i}=-70$ mV, $\gamma_{i}=10$ S/F for all the neurons and, for all the synapses, $\gamma^\mathrm{s}_{ij}=10$ S/F, $E_{\mathrm{syn},ij}=0$ mV, $a_{r,ij}=5$, $a_{d,ij}=5$, $\beta_{ij}=125\, \mathrm{V}^{-1}$. $V_{\mathrm{th},ij}=V_{\mathrm{eq},ij}$ for all the edges as in~\cite{Kunert2014}, except for edge $(\alpha,\beta)$, for which we set $V_{\mathrm{th},\alpha\beta}=-10$ mV.
For the second example, the additional inhibitory edge $3\vdash 1$ has $\gamma^\mathrm{s}_{31}=20$ S/F, $E_{\mathrm{syn},31}=-70$ mV, and $a_{r,ij}=1$, while all the other parameters of the synapse are equal to the ones above.

All the convolutions are performed using the Gregory integration scheme (of order 8), adapted from Ref.~\cite{Nessi}.